\documentclass[pra,twocolumn,showpacs, superscriptaddress,
groupeaddress,preprintnumbers,amsmath,amssymb]{revtex4}
\usepackage{graphicx}

\newcommand{\beq}{\begin{equation}}
\newcommand{\eeq}{\end{equation}}
\newcommand{\bea}{\begin{eqnarray}}
\newcommand{\eea}{\end{eqnarray}}
\newcommand{\ba}{\begin{array}}
\newcommand{\ea}{\end{array}}
\newcommand{\bc}{\begin{center}}
\newcommand{\ec}{\end{center}}

\newcommand{\commentout}[1]{{}}
\newcommand{\half}{\hbox{$1\over2$}}

\newcommand{\adag}{a^\dagger}
\newcommand{\bdag}{b^\dagger}

\newcommand{\bk}{{\bf k}}

\newcommand{\eq}[1]{(\ref{#1})}
\newcommand{\bml}{\begin{subequations}}
\newcommand{\eml}{\end{subequations}}
\newcommand{\vol}[1]{{\bf #1}}
\newcommand{\comment}[1]{{}}

\begin{document}
\title{Theory of Combined Photoassociation and Feshbach Resonances in a Bose-Einstein Condensate}

\author{Matt Mackie}
\affiliation {Department of Physics, Temple University, Philadelphia, PA 19122}
\affiliation{Department of Physics, University of Connecticut, Storrs, CT 06268}
\author{Catherine DeBrosse}
\affiliation {Department of Physics, Temple University, Philadelphia, PA 19122}
\affiliation {Department of Biology, Temple University, Philadelphia, PA 19122}
\date{\today}

\begin{abstract}
We model combined photoassociation and Feshbach resonances in a Bose-Einstein condensate, where the shared dissociation continuum allows for quantum interference in losses from the condensate, as well as a dispersive-like shift of resonance. A simple analytical model, based on the limit of weakly bound molecules, agrees well with numerical experiments that explicitly include dissociation to noncondensate modes. For a resonant laser and an off-resonant magnetic field, constructive interference enables saturation of the photoassociation rate at user-friendly intensities, at a value set by the interparticle distance. This rate limit is larger for smaller condensate densities and, near the Feshbach resonance, approaches the rate limit for magnetoassociation alone. Also, we find agreement with the unitary limit--set by the condensate size--only for a limited range of near-resonant magnetic fields. Finally, for a resonant magnetic field and an off-resonant laser, magnetoassociation displays similar quantum interference and a dispersive-like shift. Unlike photoassociation, interference and the fieldshift in resonant magnetoassociation is tunable with both laser intensity and detuning. Also, the dispersive-like shift of the Feshbach resonance depends on the size of the Feshbach molecule, and is a signature of non-universal physics in a strongly interacting system.
\end{abstract}


\maketitle


\section{Introduction}

Photoassociation occurs when an ultracold pair of atoms absorbs a photon and jumps from the two-atom continuum to a bound molecular state~\cite{LET93,BAN95}. Similarly, magnetoassociation occurs when one atom in an ultracold pair spin flips in the presence of a magnetic field tuned nearby a Feshbach resonance~\cite{INO98,STW76}. The two theories are formally identical, and are referred to in general as coherent association. If the atoms are quantum degenerate, then coherent association can be highly efficient, due to large phase space density~\cite{BUR96,HOD05}.

Pioneering experiments~\cite{WYN00} with Raman photoassociation of a Bose-Einstein condensate demonstrated highly efficient atom-molecule conversion that was just on the verge~\cite{KOS00} of coherence. Subsequent experiments~\cite{MCK02,PRO03} focused on the strongly interacting regime and the associated rate limit on atom-molecule conversion, as well as the free-bound lightshift. The rate limit for converting atoms into molecules is set by either unitarity~\cite{BOH99}, i.e., the size of the condensate, or rogue dissociation~\cite{JAV02,KOS00,NAI08}, i.e., the interparticle distance, and the lighthshift is predictedly red and proportional to laser intensity~\cite{FED96,JAV98}. Whereas the observed lightshift in Bose-condensed Na~\cite{MCK02} and ultracold $^7$Li~\cite{PRO03} was consistent with theory~\cite{FED96,JAV98} and previous non-degenerate experiments~\cite{JON97}, the rate limit could not be reached in Na and, although it was reached in $^7$Li, the absence of Bose condensation precluded a direct comparison between the rogue and unitary limits. Rate limit aside, an optically-tuned scattering length~\cite{FED96,MAC01} has been observed in an $^{87}$Rb~\cite{THE04} condensate, along with observation~\cite{WIN05} of the two photon atom-molecule dark state~\cite{VAR97,JAV98}.

On the other hand, the Feshbach resonance has garnered exceptional interest. Early experiments focused on tuning the $s$-wave scattering length in Na~\cite{INO98}, followed by the formation of a Bose-Einstein condensate of otherwise incondensable $^{85}$Rb~\cite{COR00} and $^{133}$Cs~\cite{WEB02}.  Later experiments with $^{85}$Rb demonstrated the  ``Bosenova'' collapse of a Feshbach-resonant condensate~\cite{DON01}, a counter-intuitive decrease in condensate losses for an increase in time spent near resonance~\cite{CLA02,MAC02}, as well as coherent oscillations between the Bosenova burst and its remnant core~\cite{DON02,MAC02,KOK02}. Soon after, magnetoassociation led to the molecular condensate milestone~\cite{KXU03} and, in combination with laser transfer of the vibrationally-excited Feshbach population~\cite{KOK01}, the production of a quantum degenerate gas of absolute ground state molecules~\cite{NI08}. The very latest results report that the Feshbach resonance has enabled exothermic chemical reactions in a quantum degenerate gas of dipolar molecules~\cite{OSP09}.

Meanwhile, a combination of the two resonances has also been investigated. One of the first experiments with Feshbach-resonant scattering lengths in $^{85}$Rb used a photoassociation laser as a probe~\cite{COU98}, thereby observing enhancement of the conversion from atoms to molecules, and later experiments in $^{133}$Cs observed supression~\cite{VUL99}. Supression and enhancement were then observed together in $^{133}$Cs~\cite{LAB03}, an investigation that also included a successful numerical model based on photoassociation with a Feshbach-tunable scattering length. Most recently, the Feshbach resonance was observed to vary the photoassociation rate constant by over four decades, and to anomalously shift the position of laser resonance to the blue end of the photoassociation spectrum for near-resonant magnetic fields~\cite{JUN08,MAC08}. Finally, the most recent theory predicts strongly enhanced production of dipolar molecules in the electro-vibrational ground state~\cite{PEL08}, as well as an Autler-Townes-like splitting of the combined resonance lineshape for a high intensty photoassociation laser~\cite{DEB09}. These models are analogous to laser-induced autoionization~\cite{FAN61}, and laser-enhanced photoassociation~\cite{MAC09}.

The purpose of this Article is to develop the details of the matter optics theory of combined photoassociation and Feshbach resonances. For a resonant laser and an off-resonant magnetic field, our simple analytical model~\cite{MAC08} for the rate constant and continuum shift of laser resonance agrees well with numerical experiments, with the main result indicating that losses are stronger for a more dilute gas. Next, we connect the present results to collisional models, illustrating how the rate constant and lightshift have a node at the magnetic field value where the Feshbach-resonant scattering length matches the semi-classical size of the photoassociation molecule. We also show that, for combined resonances, the rogue limit for atom-molecule conversion agrees with the unitary limit only over a limited magnetic field range near the Feshbach resonance. While these results are independent of the laser parameters, we find that, for a resonant magnetic field and an off-resonant laser, the quantum interference and fieldshift depend on both the laser intensity and detuning. Moreover, in this case the nodes in the rate constant and lightshift occur at difference intensity/detuning position, and the fieldshift depends on the size of the Feshbach molecule.

This Article is outlined as follows. In Sec.~\ref{MODEL}, we first introduce the few-level diagrams, Hamiltonian, and corresponding nonlinear Schr\"odinger equation for the combined resonance system. In Sec.~\ref{WBM}, we develop an analytical solution to the nonlinear Schrodinger equation based on the assumption of weakly bound molecules. For one or the other of the molecular states is tuned off resonance, Sec.~\ref{OFF} derives the equations of motion for the atomic and molecular probabilities, as well as the atom-molecule coherence which, in turn, leads to the rate equation for losses from the condensate. In Sec.~\ref{NUMERICS}, we report an algebraic solution to the nonlinear Schr\"odinger equation that can be implemented numerically at little computational cost, and describe our procedure for numerical experiments. Sec.~\ref{EXPLICIT} presents the parameters for $^7$Li. Results are presented in Sec.~\ref{RESULTS}. Resonant photoassociation is detailed in Sec.~\ref{ONPA}, including the comparison with numerical experiments, the connection to collision models, and the comparsion to unitarity. Resonant magnetoassociation is covered in Sec.~\ref{ONMA}, with a focus on the role of the size of the Feshbach molecule. Conclusions are given in Sec.~\ref{CONC}.

\section{Model}
\label{MODEL}

\begin{figure}[t]
\centering
\includegraphics[width=8.5cm]{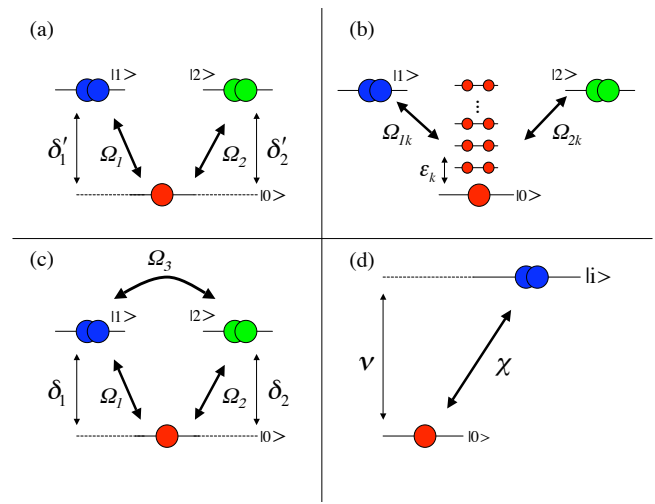}
\caption{(color online)~Few-level scheme for a condensate tuned nearby a combined photoassociation and Feshbach resonance. (a)~Basic three-level scheme, where a photoassociation and Feshbach resonance couple the atomic condensate $|0\rangle$ and molecular the condensates $|1\rangle$ and $|2\rangle$, respectively. (b)~A quasicontinuum accounts for dissociation to noncondensate levels. (c)~Eliminating the noncondensate levels leads to an effective $V$-system, where the virtual continuum couples the two molecular states, and where the detunings $\delta_i$ include the free-bound redshift. (d)~When the system is far from one resonance, magnetic or laser, the off-resonant molecular state can also be eliminated, leaving an effective two-level system, where the detuning $\nu$ includes an anomalous Stark-shift.}
\label{FEWL2}
\end{figure}

Consider $N$ atoms that have Bose condensed into, say, the zero momentum plane-wave ($\hbar\bk=0$)
state $|0\rangle$. Photoassociation and the Feshbach resonance then couple atoms in the state $|0\rangle$ to diatomic molecules of zero momentum in the states $|1\rangle$ and $|2\rangle$, respectively. As per Fig.~\ref{FEWL2}(a), this is the $V$-system familiar from few-level laser spectroscopy. Annihilation of an atom (molecule) of mass $m$ ($M=2m$) from the atomic ($i$th molecular) condensate is represented by the second-quantized operator $a_0\equiv a$ ($b_i$). This theory is the simplest, and molecules dissociate only back to the level $|0\rangle$. To be more complete, molecular dissociation to noncondensate levels should also be included [Fig.~\ref{FEWL2}(b)]. These levels must be considered because a condensate molecule need not dissociate back to the atomic condensate, but may just as well create a pair of atoms with equal-and-opposite momentum, since total momentum is conserved. So-called rogue
\cite{KOS00,JAV02}, or unwanted \cite{GOR01}, dissociation to noncondensate
modes therefore introduces the operators $a_{\pm\bk}$.

The second quantized Hamiltionian for this system is $\hat{H}=\hat{H}_0+\hat{H}_1$, where the condensate Hamiltonian is
\bml
\beq
\frac{\hat{H}_0}{\hbar}=\sum_{i=1,2}\left[\delta_i'\bdag_i b_i -\kappa_i(\bdag_iaa+\adag\adag b)\right],
\label{H_CON}
\eeq
and the noncondensate Hamiltonian is
\beq
\frac{\hat{H}_1}{\hbar}=\half\sum_\bk\varepsilon_\bk\adag_\bk a_\bk
  -\sum_i\kappa_i\sum_{\bk\neq0}
    f_{i\bk}\left[ \bdag_i a_{-\bk}a_\bk +\adag_\bk\adag_{-\bk}b_i \right]
 \label{H_NONCON}
\eeq
\label{HAM}
\eml
Here the bare photoassociation (Feshbach) detuning is $\delta_1'$ ($\delta_2'$), the atom-molecule coupling is $\kappa_1$ ($\kappa_2$), the kinetic energy of an atom pair of reduced mass $m_r=m/2$ is $\hbar\varepsilon_\bk=\hbar^2k^2/2m_r$, and the momentum dependence of the coupling to the noncondensate modes is given by $f_{1\bk}$ ($f_{2\bk}$) with $f_{i\bf 0}=1$. Spontaneous decay of the photoassociation molecule has been neglected for now, whereas spontaneous decay of the Feshbach molecule~\cite{KOE05} is neglected once and for all. 

To obtain mean-field equations, the Heisenberg equation for a given
operator, $i\hbar\dot{x}=[x,\hat{H}]$, is derived from the Hamiltonian~\eq{HAM}, and all operators are subsequently declared $c$-numbers. In a minimalist model, $x$ represents either the atomic ($i$th molecular) operator $a_\bk$ ($b_i$), or the anomalous density operator $A_\bk=a_\bk a_{-\bk}$, where $A_\bk$ arises from rogue
dissociation to noncondensate atom pairs of equal-and-opposite momentum. Converting summations over $\bk$ to integrals over frequency $\varepsilon$ introduces the characteristic frequency $\omega_\rho=\hbar\rho^{2/3}/2m_r$. The resulting mean-field theory is
\bml
\bea
i\dot{a} &=& -\Omega_1 a^* b_i -\Omega_2 a^* b_2, 
\\
i\dot{b}_i &=& \delta_i' b_i
  -\half\Omega_i a^2-\half\xi_i\!\int\!d\varepsilon\sqrt{\varepsilon}\,f_i(\varepsilon)A(\varepsilon),
\label{BDOT}
\\
i\dot{A}(\varepsilon) &=&\varepsilon A(\varepsilon)
    -\Omega_1\,f_1(\varepsilon)b_1-\Omega_2\,f_2(\varepsilon)b_2.
\label{ADOT}
\eea
\label{BOSE_EQM}
\eml
These amplitudes are of unit order, $\Omega_i=\sqrt{N}\kappa_i\propto\sqrt{\rho}$, and $\xi_i=\Omega_i/(4\pi^2\omega_\rho^{3/2})$. Finally, the continuum shape is related to the size of the underlying molecule and, since there are two molecules, we allow for different length scales with the Gaussian $f_i=\exp[-(\varepsilon/\beta_i)^2/2]$, where $\beta_i=\hbar/(2m_rL_i^2)$ with $L_i$ the semi-classical size of the $i$-th molecular state. Here we choose a Gaussian over a Lorentzian~\cite{MAC08} to foster faster numerical convergence.

\subsection{Weakly-Bound Molecules}
\label{WBM}

Previously, it was shown that the shared dissociation continuum can act like a virtual state~\cite{MAC08,MAC09}, leading to an effective cross-coupling between the two molecular states. However, this result is based on adiabatic elimination of the anomalous amplitude which, despite capturing the essential physics, leaves something to be desired. This subsection therefore develops the cross-molecular coupling based on simple Fourier analysis.

Consider the equations of motion~\eq{BOSE_EQM} in the limit $\Omega_i=0$ with $\xi\neq0$ and $\Omega_if_i(\varepsilon)\neq0$. Fourier transforming the resulting equations and subsequently eliminating the rogue amplitude $\tilde{A}(\omega)$ gives
\bml
\bea
i\tilde{\dot b}_1(\omega)&=&\left[ \delta_1'+\Sigma_1(\omega) \right]\tilde{b}_1(\omega)
  +\half\Omega_3(\omega)\tilde{b}_2(\omega)
  \\
i\tilde{\dot b}_2(\omega)&=&\left[ \delta_2'+\Sigma_2(\omega) \right]\tilde{b}_2(\omega)
  +\half\Omega_3(\omega)\tilde{b}_1(\omega),
\eea
\label{MOL_AMP_FOURIER}
\eml
where the frequency dependent shift and cross molecular coupling are
\bml
\bea
\Sigma_i(\omega)&=&\frac{\Omega_i^2}{8\pi^2\omega_\rho^{3/2}}
  \int d\varepsilon\sqrt{\varepsilon}\,\frac{f_i^2(\varepsilon)}{\omega-\varepsilon},
\\
\tilde{\Omega}_3(\omega)&=&\frac{\Omega_1\Omega_2}{4\pi^2\omega_\rho^{3/2}}
  \int d\varepsilon\sqrt{\varepsilon}\,\frac{f_1(\varepsilon)f_2(\varepsilon)}{\omega-\varepsilon}.
\eea
\eml
In the limit of weakly bound molecules, $\sigma_{0i}=\lim_{\omega\rightarrow0}\Re[\Sigma_i(\omega)]$, $\gamma_i/2=\lim_{\omega\rightarrow0}\Im[\Sigma_i(\omega)]$, and $\Omega_3=\lim_{\omega\rightarrow0}\Re[\tilde\Omega_3(\omega)]$. Taking the inverse transform then yields
\bml
\bea
i\dot{b}_1&=&(\delta_1-i\gamma_1/2)b_1-\half\Omega_3 b_2
\\
i\dot{b}_2&=&(\delta_2-i\gamma_2/2)b_2-\half\Omega_3 b_1
\eea
\label{CROSS_CPLD}
\eml
where the continuum-shifted (renormalized) detuning is now $\delta_i=\delta_i'-\sigma_{0i}$. 

The equations of motion~\eq{CROSS_CPLD} imply that, to lowest order, the non-condensate Hamiltonian~\eq{H_NONCON} is
\beq
\frac{\hat{H}_1}{\hbar}\approx-\half\Omega_3(\bdag_1b_2+\bdag_2 b_1),
\label{H_CROSS}
\eeq
and the full mean-field theory ($\Omega_i\neq0$) is, to lowest order, given as
\bml
\bea
i\dot{a} &=& -\Omega_1 a^* b_1 -\Omega_2 a^* b_2, 
\\
i\dot{b}_1 &=& \delta_1 b_1
  -\half\Omega_1 a^2-\half\Omega_3b_2,
\\
i\dot{b}_2 &=& \delta_2 b_2
  -\half\Omega_2 a^2-\half\Omega_3b_1.
\label{B2DOT}
\eea
\label{CROSS_EQM}
\eml
As per Eqs.~\eq{H_CROSS}~and~\eq{CROSS_EQM}, the shared dissociation continuum of Fig.~\ref{FEWL2}(b) can be approximated by an effective cross coupling between the molecules, illustrated in Fig.~\ref{FEWL2}(c).

Now, any time a continuum is coupled to a bound state, the eigenfrequency of the bound state is necessarily shifted. This shift is to the red, is well understood~\cite{FED96,JAV98,JON97,MCK02,PRO03}, and has already been included in the renormalized detuning, $\sigma_{0i}=\Re[\Sigma_i(0)]<0$. However, the continuum here is coupled to a pair of bound states, and the effective cross-molecular coupling means that the Feshbach bound state will shift the eigenfrequency of the photoassociation resonance, and vice versa. Returning to the Fourier expressions for the cross-coupled molecular amplitudes [Eqs.~\eq{MOL_AMP_FOURIER}], and simplifying the transform of the derivatives, yields the lowest order result for the shift of the photoassociation and magnetoassociation resonances, respectively,
\bml
\bea
\sigma_1&=&\sigma_{01}
  +\half\left[ \delta_2+\delta_1\pm\sqrt{(\delta_2-\delta_1)^2+\Omega_3^2} \right],
\nonumber\\
\\
\sigma_2&=&\sigma_{02}
  +\half\left[ \delta_1+\delta_2\pm\sqrt{(\delta_1-\delta_2)^2+\Omega_3^2} \right].
  \nonumber\\
\eea
\label{SHIFTS}
\eml
Here the positive (negative) root gives the shift above (below) resonance. In contrast to the shift for a single bound state, the bound Feshbach (photoassociation) molecular state, along with the shared dissociation continuum, leads to a net shift of the photoassociation (magnetoassociation) resonance that turns from red to blue for $\delta_2\agt-\Omega_3^2/|\sigma_{01}|$ ($\delta_1\agt-\Omega_3^2/|\sigma_{02}|$), where we have assumed $(\sigma_{0i}/\Omega_3)^2\ll1$. In particular, the redshift depends on the size of the underlying molecule, and therefore the position of the node in the cross-molecular shift does too.

\subsection{Off-Resonant Systems}
\label{OFF}

We now detail a simple analytical model for the cross-coupled system where one of the fields, magnetic or laser, is tuned far from resonance. In this case, the off-resonant molecular amplitude can be adiabatically eliminated from the mean-field theory, leading to the effective two-level system shown in Fig.~\ref{FEWL2}(d). For the $i$th ($j$th) molecular state tuned onto (off) resonance, $i\dot{b}_i/\delta_i\approx0$ and the mean-field theory~\eq{CROSS_EQM} becomes
\bml
\bea
i\dot{a}&=&-\chi a^*b_i,
\\
i\dot{b}_i&=&(\nu-i\Gamma/2) b_i-\half\chi a^2,
\eea
\label{TWOL}
where the effective atom-molecule coupling, the Stark-shifted detuning, and the effective damping, respectively, are given as
\bea
\chi&=&\Omega_i+2{\cal L}_j\delta_j\Omega_j/\Omega_3,
\\
\nu&=&\delta_i-{\cal L}_j\delta_j
\\
\Gamma&=&\Gamma_i+{\cal L}_j\Gamma_j,
\eea
\label{RES_PARAMS}
\eml
where ${\cal L}_j=\Omega_3^2/4|\tilde\delta_j|^2$. Note that this lighthsift shift agrees with Eqs.~\eq{SHIFTS} in the off-resonant limit $\delta_i\gg\delta_j,\Omega_3$. Here the decay rate of the photoassociation (Feshbach) molecule is $\Gamma_1=\Gamma_s+\gamma_1$ ($\Gamma_2=\gamma_2$), where $\Gamma_s$ is the spontaneous electronic decay rate of the photoassociation bound state, spontaneous decay of the Feshbach state has been ignored, and the dissociative decay rate is $\gamma_i=\Im[\Sigma_i(0)]$. 

We now follow Ref.~\cite{KOS00}, and derive a rate equation for the atomic and molecular probabilities, $P_0=a^*a$ and $P_i=2b_i^*b_i$, as well as the atom-molecule ``coherence" $C_{0i}=aab_i^*$. Making extensive use of the product rule, e.g., $i\dot C_{0i}=iaa\dot b_i^*+2ia\dot ab_i^*$, we have
\bml
\bea
i\dot P_0 &=& \chi(C_{0i}-C_{0i}^*), \\
i\dot P_i &=& -i\Gamma P_i -\chi(C_{0i}-C_{0i}^*), \\
i\dot C_{0i} &=& -(\nu+i\Gamma/2)C_{0i}
  +\half \chi P_0(P_0-2P_i).
\eea
\label{GEN_RATE}
\eml
Solving Eqs.~\eq{GEN_RATE} in the reservoir approximation, $P_0\sim1$ and $P_i\ll1$, and for an adiabatic coherence $\:\dot{C}_{0i}\approx0$, the rate equation for losses from the atomic condensate is
\bml
\beq
\dot P_0 = \rho K P_0^2,
\label{RATE_EQN}
\eeq
where the rate constant for condensate losses is
\beq
\rho K = \frac{1}{2}\,\frac{\chi^2\Gamma}{\nu^2+\Gamma^2/4}\,.
\label{RATE_EQN}
\eeq
\eml

In addition to the anomalous lightshift, we realize that, as a function of the off-resonant detuning, $\delta_j$, the resonant rate constant $K$ is suppressed below the rate for the on-resonant field alone, $\chi<\Omega_i$, essentially vanishes for $\chi\approx0$, and is enhanced above the rate for the on-resonant field alone for $\chi>\Omega_i$. Borrowing intuition from quantum optics, this suppression and enhancement is due to quantum interference between direct resonant association and association via the off-resonant molecular state.

\subsection{Numerical Experiments}
\label{NUMERICS}

We also develop an exact numerical solution of Eqs.~\eq{BOSE_EQM}, which is valid even for dual resonance. Off-resonant fields lead to stiff equations, so the usual numerical routines require long run times. On the other hand, most stiff routines are based on matrix approaches with run times that scale quadratically with the number of quasicontinuum states, leading to another impasse. 

Hence, we write the equations of motion~\eq{BOSE_EQM} in the matrix form 
$i\dot\psi=H\psi$, where semi-classical Hamiltonian and wavefuntion are defined by
\begin{widetext}
\bea
H\psi&=&
\left(
\begin{array}{cccccc}
\delta_1 & 0 & -\frac{\Omega_1}{\sqrt{2}}\,a 
  & -\frac{\xi_1}{\sqrt{2}}\sqrt{\varepsilon_1}f_1(\varepsilon_1)d\varepsilon 
    & \cdots 
      & -\frac{\xi_1}{\sqrt{2}}\sqrt{\varepsilon_{N_Q}}f_1(\varepsilon_{N_Q})d\varepsilon 
\\
0 & \delta_2 & -\frac{\Omega_2}{\sqrt{2}}\,a 
  & -\frac{\xi_2}{\sqrt{2}}\sqrt{\varepsilon_1}f_2(\varepsilon_1)d\varepsilon 
    & \cdots 
      & -\frac{\xi_2}{\sqrt{2}}\sqrt{\varepsilon_{N_Q}}f_2(\varepsilon_{N_Q})d\varepsilon 
\\
-\frac{\Omega_1}{\sqrt{2}}\,a^* & -\frac{\Omega_2}{\sqrt{2}}\,a^* & 0 & 0 
  & \cdots & 0 
\\
-\frac{\Omega_1}{\sqrt{2}}\,f_1(\varepsilon_1) 
  & -\frac{\Omega_2}{\sqrt{2}}\,f_2(\varepsilon_1) & 0 & \varepsilon_1 
    & \ddots  & \vdots
\\
\vdots & \vdots & \vdots & \ddots & \ddots  & 0
\\
-\frac{\Omega_1}{\sqrt{2}}\,f_1(\varepsilon_{N_Q}) 
  & -\frac{\Omega_2}{\sqrt{2}}\,f_2(\varepsilon_{N_Q}) & 0 & \cdots & 0
    & \varepsilon_{N_Q}
\end{array}
\right)
\left(
\begin{array}{c}
b_1 \\ b_2 \\ a \\ A(\varepsilon_1) \\ \vdots \\ A(\varepsilon_{N_Q})
\end{array}
\right).
\eea
\end{widetext}
The molecular amplitudes have been scaled by $\sqrt{2}$ so that the molecular probability is $P_i=|b_i|^2$, and the integration over the noncondensate modes has been broken into a simple summation. Here we see that $H$ is characteristically sparse, i.e., only the first two rows, the first two columns, and the diagonal contain non-zero elements. Hence, the double-resonance system can be solved using an adapted Crank-Nicholson routine~\cite{MAC99}, with run times that scale linearly with the number of states. In particular, given the solution $\psi_0=\psi(t)$ at time $t$, the solution $\psi=\psi(t+dt)$ at time $t+dt$ is given as
\beq
\psi=\frac{1-i(dt/2)H}{1+i(dt/2)H}\,\psi_0.
\label{CRANK}
\eeq
Defining $M_\pm=1\pm(dt/2)H$, Eq.~\eq{CRANK} can be written $\psi=M_-M_+^{-1}\psi_0$ so that, for $\phi$ the solution to $M_+\phi=\psi_0$, then the desired solution is simply $\psi=M_-\phi$. Nevertheless, $H=H(\psi)$ means that coherent association is inherently nonlinear, and a predictor-corrector approach is therefore taken: $H[\psi_0]$ determines the solution $\bar\psi$ from the initial solution $\psi_0$, and $H[\bar\psi]$, in turn, determines the sought after solution $\psi$.

Finally, these are numerical experiments in the truest sense of the words. First, the numerical rate constant for photoassociation is defined by $\rho K\tau=1$, where $\tau$ is the time required for the condensate fraction to fall to $|a|^2=1/e\approx0.3678$. Also, the numerical lightshift is determined by minimizing $\tau$ with respect to the laser detuning. In other words, we set the laser detuning to an inital guess $\delta_1)_{\rm guess}\sim\Gamma_s$, determine $\tau$; reset the laser detuning, update $\tau$; repeating until $\tau$ stops decreasing and begins to increase, which defines the minimum.

\subsection{Explicit Parameters}
\label{EXPLICIT}

We now determine typical parameters for each resonance. Starting with photoassociation, we set $i=1$ and ${\cal L}_2=0$ in Eq.~\eq{RATE_EQN}. In the low-intensity limit, spontaneous electronic decay dominates dissociation, $\gamma_1\ll\Gamma_s$, since $\gamma_1\propto\Omega_1^2\propto I$, so that the effective loss rate is $\Gamma\approx\Gamma_s$. Equation~\eq{RATE_EQN} for the rate constant then becomes $\rho K_0=2\Omega_2^2/\Gamma_S$. Defining $K_0=K'_0I$ and $\Omega_1=\bar\Omega_1\sqrt{I}$, the photoassociation Rabi coupling is determined by 
$\bar\Omega_1=\sqrt{\rho K'_0\Gamma_s/2}$. In the high intensity limit, $\gamma_1\gg\Gamma_s$ and decay is dominated by dissociation, $\Gamma\approx\gamma_1$, so that the saturated rate constant is defined by $\rho K_\infty=2\Omega_1^2/\gamma_1$, so that  $\gamma_1/I=K_0'\Gamma_s/K_\infty$. Accordingly, the saturation intensity is defined by $I_0=\Gamma_S/(\gamma_1/I)$, given explicitly as $I_0=K_\infty/K_0'$. Also, the characteristic length scale for saturated photoassociation is 
$\ell_1=mK_\infty/(8\pi\hbar)$. Lastly, the width of the photodissociation continuum is obtained from the lightshift per unit intensity $\sigma_{01}=\sigma'_{01}I$, so that $\beta_1=\hbar/(m L_1^2)$ and
$L_1=mK'_\infty\Gamma_s/(4\pi^2 \hbar\sigma'_{01})$.

Turning to magnetoassociation, experiments usually measure the resonant scattering length defined by $a_{\rm res}=a_{\rm bk}\Delta_B/(B-B_0)$, where $\Delta_B$ is the width and $B_0$ is the magnetic field position of the Feshbach resonance, and where $a_{\rm bk}$ is the zero-field scattering length. In the present model, the resonant scattering length is defined from Eqs.~\eq{TWOL} for $i=2$ and ${\cal L}_1=0$, combined with the limit $\delta_2\gg\gamma_2$, so that $a_{\rm res}=-\Omega_2^2/\delta_2$. Here we have assumed the continuum redshift is included in the definition of $B_0$. The Feshbach Rabi coupling is then $\Omega_2=\sqrt{8\pi\rho|a_{\rm bk}|\Delta_B\Delta_\mu/m}$, and the detuning is $\delta_2=(B-B_0)\Delta_\mu/\hbar$, where $\Delta_\mu$ is the difference in magnetic moments between the Feshbach molecule and the free-atom pair. 

Focusing on $^7$Li~\cite{PRO03}, typical photoassociation parameters are then $\Gamma_s=12\times2\pi$~MHz, $K_0'=7.9\times10^{-10}$~(cm$^3$/s)/(W/cm$^2$), $K_\infty=2.2\times10^{-8}$~cm$^3$/s, $\Sigma_1'=1.7\times2\pi$~MHz/(W/cm$^2$), and $\rho=4\times10^{12}$~cm$^{-3}$. The atom-molecule Rabi coupling is then $\bar\Omega_1\sqrt{I_0}=290\times2\pi$~kHz, and the photoassociation saturation intensity is $I_0=28$~W/cm$^2$. From the formal definition of the photodissociation rate, $2\gamma_1=\pi\xi\Omega_1\sqrt{\omega_1}$ with $\omega_1=\hbar/(m\ell_1^2)$, the photodissociation length scale is $\ell_1=\lambdabar/1.11$, where $2\pi\lambdabar=671$~nm is the wavelength of the photoassociating light. Lastly, the width of the dissociation continuum is $\beta_1=29\times2\pi$~MHz, so that $L_1=133a_0$ is consistent with the Condon radius of the photoassociation molecule~\cite{PRO03,JUN08} ($a_0$ is the Bohr radius). Note that the cleaning up of dimensionless constants of order unity have led to slightly different values from Ref.~\cite{MAC08}.

In $^7$Li ~\cite{JUN08}, a Feshbach resonance is located at $B_0=736$~G, the product of the zero-field scattering length and the resonance width is $|a_{\rm bk}|\Delta_B=1.6$~nm$\cdot$G, and the difference in magnetic moments between the Feshbach molecule and the free-atom pair is $\Delta_\mu=2\mu_0$ (the Bohr magneton is $\mu_0$). The Feshbach atom-molecule coupling is then $\Omega_2=127\times2\pi$~kHz. The sole unfixed parameter is the kinetic energy of the magnetodissociated pair, $\hbar\omega_2=q_2^2/2m_r$. For a zero-temperature homogeneous system of point-like Feshbach molecules, the reasonable ansatz for the length scale is the interparticle distance, so that the uncertainty principle gives $q_2= \hbar\rho^{1/3}$.

\section{Results}
\label{RESULTS}

This section presents our analytical and numerical results for resonant lasers and off-resonant magnetic fields, and vice versa. Off-resonant detunings are restricted to below threshold (negative) values, since condesates are unstable for negative resonant scattering lengths.

\subsection{Resonant Photoassociation}
\label{ONPA}

\subsubsection{Comparison to Numerical Experiments}

The purpose of this section is a detailed comparison between our previously reported analytical model~\cite{MAC08} with numerical experiments. Away from the Feshbach resonance and on lightshifted resonance ($\nu=0$), the rate constant is
\bea
\rho K&=&
  \frac{\Omega_1^2\left[1+\eta\left({\displaystyle \frac{1}{\rho^{1/3}L_1}}\right)
    \left({\displaystyle\frac{\Omega_2^2}{8\pi\omega_\rho\delta_2}}\right)\right]^2}
  {\Gamma_1+\left[\eta
    \left({\displaystyle\frac{1}{\rho^{1/3}L_1}}\right)
      \left({\displaystyle\frac{\Omega_2^2}{8\pi\omega_\rho\delta_2}}\right)
        \left({\displaystyle\frac{\Omega_1}{\Omega_2}}\right)\right]^2
          \gamma_2}\,,\nonumber\\
\label{RES_RATE_CON}
\eea
where $\eta\sim1$ is a numerical factor leftover in $\Omega_3$ leftover from the integration of the Gaussian continuum shapes, and we have assumed point-like Feshbach molecules $L_2\ll L_1$ (when need for numerics, we use $L_2=40a_0$). For increasing laser intensity, the saturated rate constant is
\bml
\beq
\rho K=
  \left[1+\frac{1}{\eta}\,\rho^{1/3}L_1\,
    \left(\frac{8\pi\omega_\rho\delta_2}{\Omega_2^2}\right)\right]^2
  \frac{\Omega_2^2}{\gamma_2}\,,
\eeq 
and the enhanced saturation intensity is
\beq
\frac{I_0'}{I_0}=\left[\frac{1}{\eta}\,(\rho^{1/3}L)
        \left(\frac{8\pi\omega_\rho\delta_2}{\Omega_2^2}\right)
      \left(\frac{\Omega_2}{\Omega_{01}}\right)\right]^2
   \frac{\Gamma_{01}}{\gamma_2}.
\eeq
Near the free-bound resonance ($\delta_2\sim0$), the rate constant saturates at the rate constant for free-bound photoassociation alone, $\rho K=\Omega_2^2/\Gamma_2$, and the saturation intensity tends to zero. On destructive interference, photoassociation ceases for a magnetic field position that depends on the size of the Feshbach molecule:
\beq
B_n=B_0-\frac{\eta}{8\pi}\,
  \frac{\hbar\Omega_2^2}{\Delta_\mu\omega_\rho}\,\rho^{1/3}L_1.
\label{B_NODE}
\eeq
Finally, the anomalous lightshift is 
\beq
\sigma_1=\sigma_{01}\left[1+\eta^2\,\left(\frac{1}{\rho^{1/3}L_1}\right)\,
  \left(\frac{\Omega_2^2}{8\pi\omega_\rho\delta_2}\right)\right],
\eeq
which has its own node at the magnetic field position
\beq
B=B_0-\frac{\eta^2}{8\pi}\,
  \frac{\hbar\Omega_2^2}{\Delta_\mu\omega_\rho}\,\rho^{1/3}L_1.
\label{B_NODE}
\eeq
\eml

\begin{figure}[h]
\centering
\includegraphics[width=8.5cm]{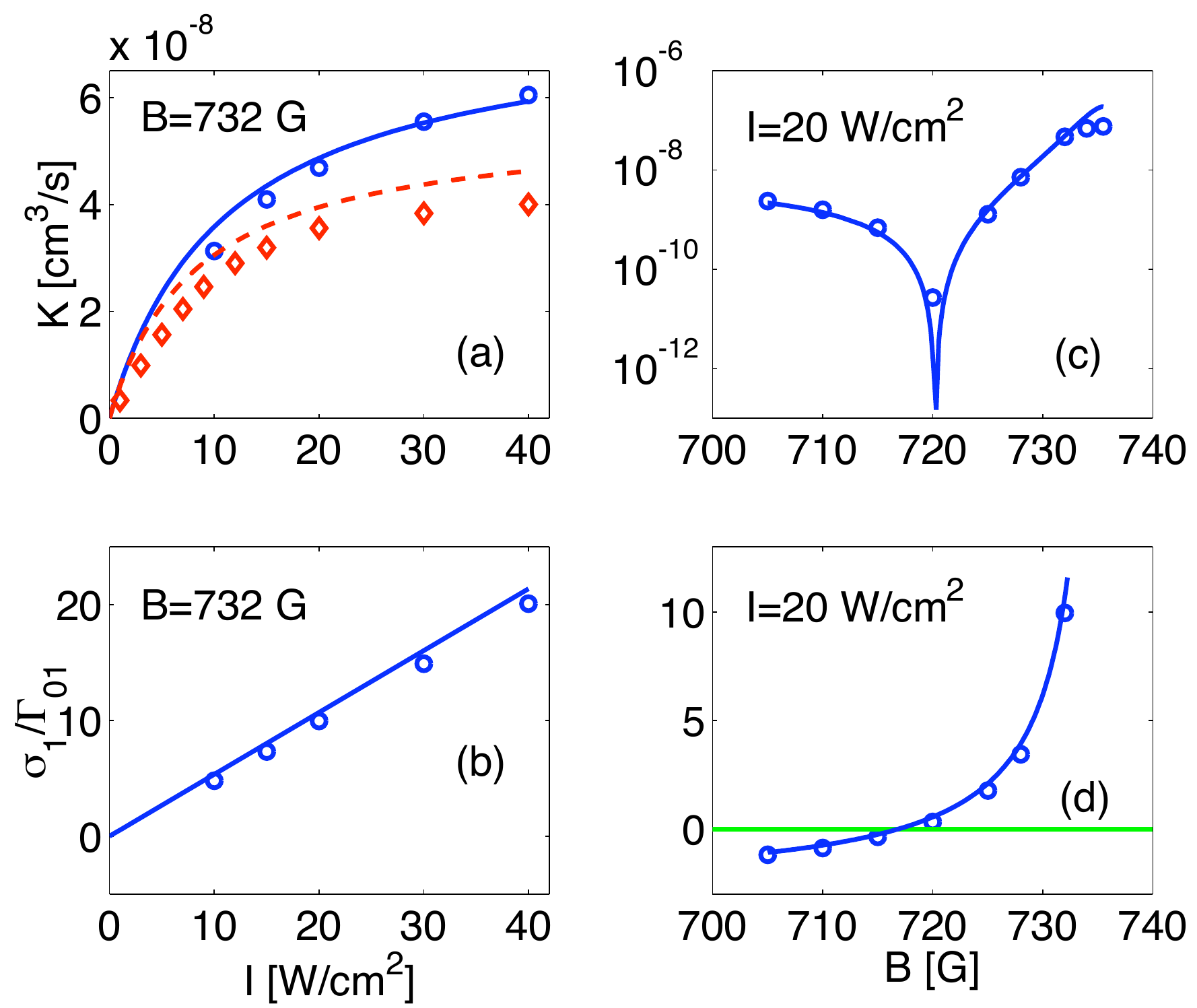}
\caption{(color online)~Rate constant and lightshift for resonant photoassociation and off-resonant magnetoassociation, as a function of laser intensity (a,b) and magnetic field (c,d). Also in panel~(a), dependence of the rate constant on density, where the blue solid (red dashed) line is the analytical result and the blue open circles (red diamonds) are the numerical results for $\rho=1\times10^{12}$~cm$^{-3}$ ($\rho=4\times10^{12}$~cm$^{-3}$).}
\label{ALL}
\end{figure}

Results are shown in Fig.~\ref{ALL}. As a function of laser intensity, the analytical and numerical rate constants agree best for low density, suggesting that the model is most reliable when rogue dissociation to noncondensate modes is dominant. As a function of magnetic field, and for fixed intensity, the agreement holds steady except near resonance, most likely because the adiabatic approximation used to eliminate the Feshbach molecular amplitude breaks down. The agreement between the analytical and numerical results for the lightshift is also good.

\subsubsection{Connection to Collision Models}

We may also make the connection to existing models~\cite{VUL99,JUN08} of photoassociation near a Feshbach resonance. For an off-resonant field, the resonant interatomic scattering length is defined $4\pi\hbar\rho a_{\rm res}/m=-\Omega_2^2/2\delta_2$. The rate constant~\eq{RES_RATE_CON} is then
\beq
\rho K=
  \frac{\Omega_1^2\left(1-\eta{\displaystyle \frac{a_r}{L}}\right)^2}
  {\Gamma_1+\left(\eta
   {\displaystyle\frac{a_r}{L}}
       {\displaystyle\frac{\Omega_1}{\Omega_2}}\right)^2
          \gamma_2}\,.
\label{RES_RATE_CON_COLL}
\eeq
Similarly, the saturated rate constant, the saturation intensity, and the lightshift are re-written as
 \bml
\bea
\rho K&=&
  \left(1-\frac{1}{\eta}\,\frac{L}{a_r}\right)^2\frac{\Omega_2^2}{\Gamma_2}
\\
\frac{I_{01}'}{I_{01}}&=&
  \left(\frac{1}{\eta}\,\frac{L}{a_r}\frac{\Omega_2}{\Omega_{01}}\right)^2
     \frac{\Gamma_1}{\Gamma_2},
\\
\sigma_1&=&\sigma_{01}\left(1-\eta^2\frac{a_r}{L}\right).
\eea
\eml

\begin{figure}[t]
\centering
\includegraphics[width=4.5cm]{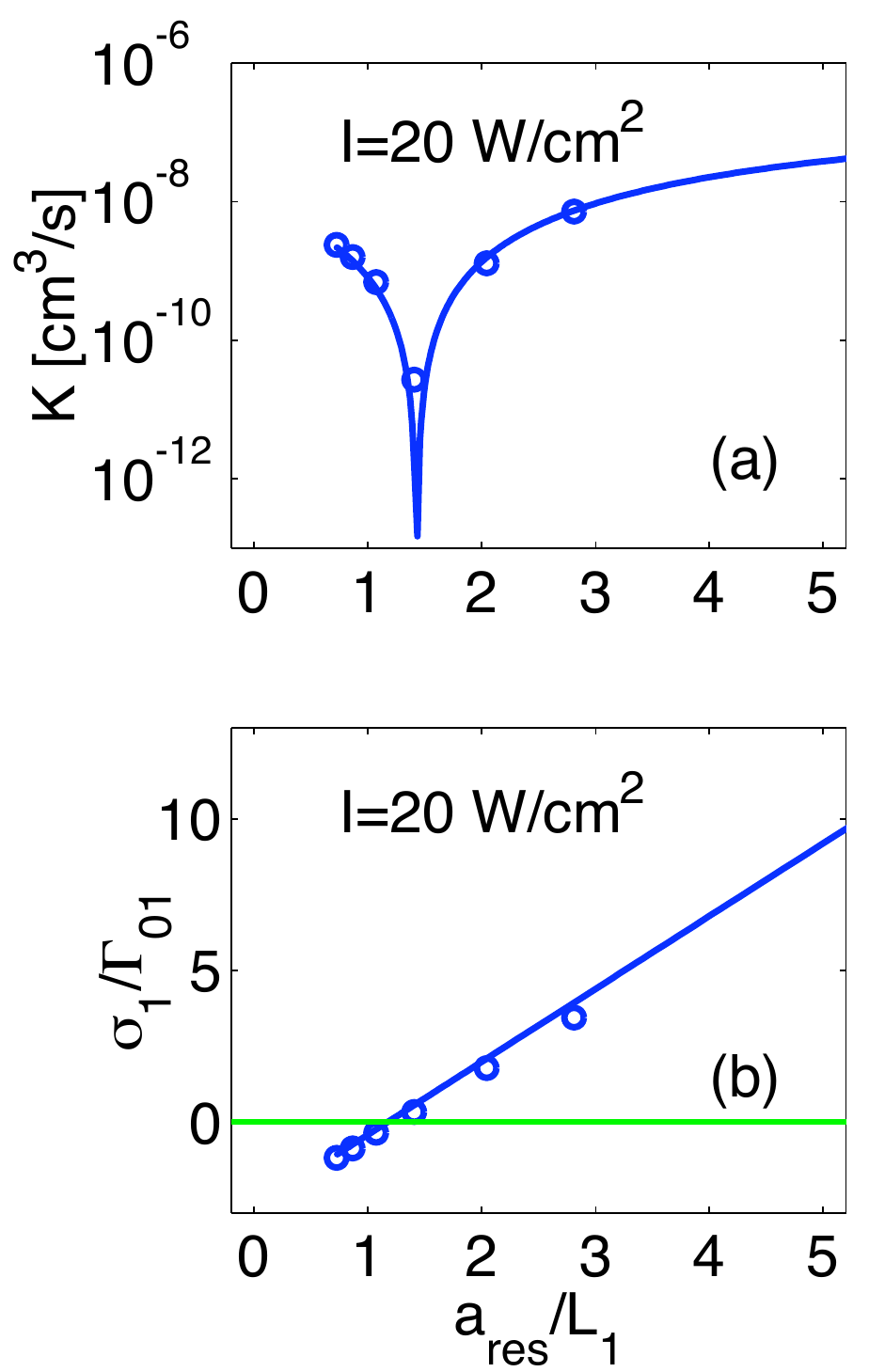}
\caption{(color online)~Rate constant and lightshift for resonant photoassociation and off-resonant magnetic fields, as a function of the resonant scattering length. Here the density is $\rho=1\times10^{12}$~cm$^{-3}$, the solid line is the analytical result, and the open circles are the numerical results.}
\label{NodeShift_vAres}
\end{figure}

Results are shown in Fig.~\ref{NodeShift_vAres}. Both analytically and numerically, the rate constant for resonant photoassociation ceases when the Feshbach-resonant scattering length matches the size of the photoassociation molecule, $a_r\sim L_1$. The node in the lighthshift occurs similarly. However, this coincidence is due to the limit where the Feshbach state is much smaller than the excited state, $L_2\ll L_1$; if the two molecular sizes are comparable (but not equal), or the Feshbach state were much larger, then the rate constant and the lightshift vanish at different magnetic field locations.

\subsubsection{Comparison to Unitarity}

Here we explicitly compare the rogue and unitary limits. The unitary limit is generally set by the de~Broglie wavelength which, for a zero-temperature Bose-Einstein condensate, is given by the size of the condensate. In the Thomas-Fermi model, the size of the condensate is determined from the trapping frequency $\nu=\{\nu_1,\nu_2,\nu_3\}$ and the condensate chemical potential $2\mu/\bar{\omega}=\sqrt[5]{(15Na_{\rm res}/a_{\rm ho})^2}$, where $a_{\rm ho}=\sqrt{\hbar/m\bar{\nu}}$ is the harmonic oscillator length scale with $\bar{\nu}=\sqrt[3]{\nu_1\nu_2\nu_3}$. Finally, $a_{\rm res}$ is the Feshbach-resonant s-wave scattering length given by $4\pi\hbar\rho a_{\rm res}/m=-\Omega_2^2/(2\delta_2$). Focusing on cigar-shaped traps, the condensate is tightly confined in the $xy$-plane and loosely confined along the $z$-direction, i.e., $\nu_1=\nu_2=\nu_r$ and $\nu_r\gg\nu_3$. The unitarity-limited rate constant is then
\bml
\beq
K_u=\frac{\hbar\Lambda_d}{m},
\label{UNITARITY}
\eeq
where the de~Broglie wave length is set by the Thomas-Fermi radius of the cigar
\beq
\Lambda_d=2R_{\rm TF}=\sqrt{\frac{8\hbar\mu}{m\nu_r^2}}.
\eeq
\eml

\begin{figure}[t]
\centering
\includegraphics[width=4.5cm]{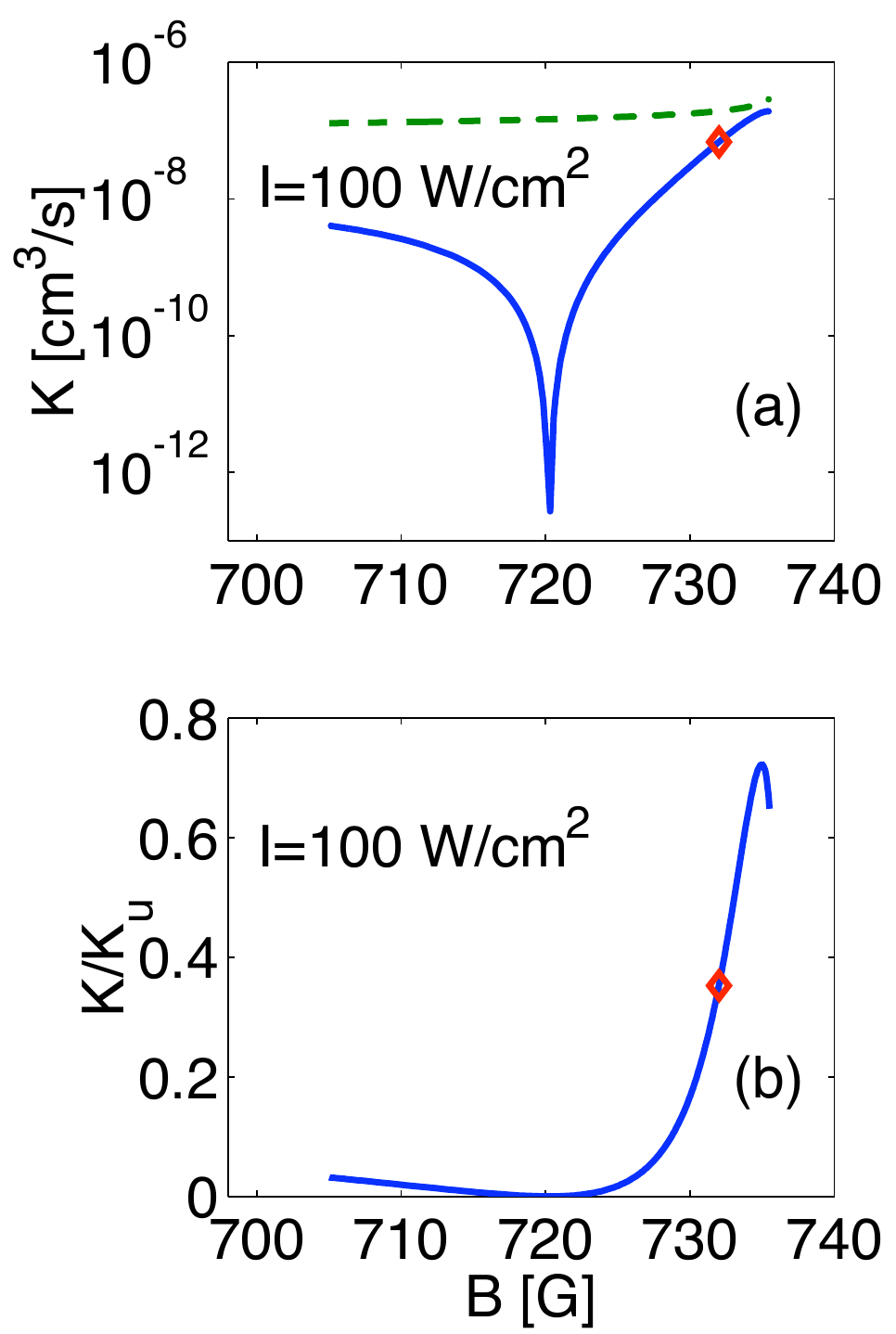}
\caption{(color online)~Unitarity vs. rogue limit for resonant photoassociation and near-resonant magnetoassociation, as a function of magnetic field. The laser intensity is set to $I=100$~W/cm$^2$ to ensure saturation absent Feshbach enhancement. (a) The rogue (unitary) rate constant is given by the solid (dashed) line. (b)~Ratio of rogue to unitary rate constants. The red diamond is not a numerical result, but simply marks $B=732$~G for comparison to Fig.~\ref{ALL}(a).}
\label{RvU}
\end{figure}

In Fig.~\ref{RvU}, we see that the rogue and unitary limits only agree reasonably for magnetic fields tuned within a small window near the Feshbach resonance, and the two limits diverge considerably as the system moves away from and onto the Feshbach resonance. It is no surprise that unitarity does not account for the node, and we attribute the discrepancy near resonance to a breakdown of both the Thomas-Fermi and rogue models as the system approaches threshold, where Feshbach molecules should be explicitly included. Far below threshold, where photoassociation alone dominates, the rogue limit is more strict than unitarity, as per Ref~\cite{JAV02}.

\subsection{Resonant Magnetoassociation}
\label{ONMA}

\begin{figure}[t]
\centering
\includegraphics[width=8.5cm]{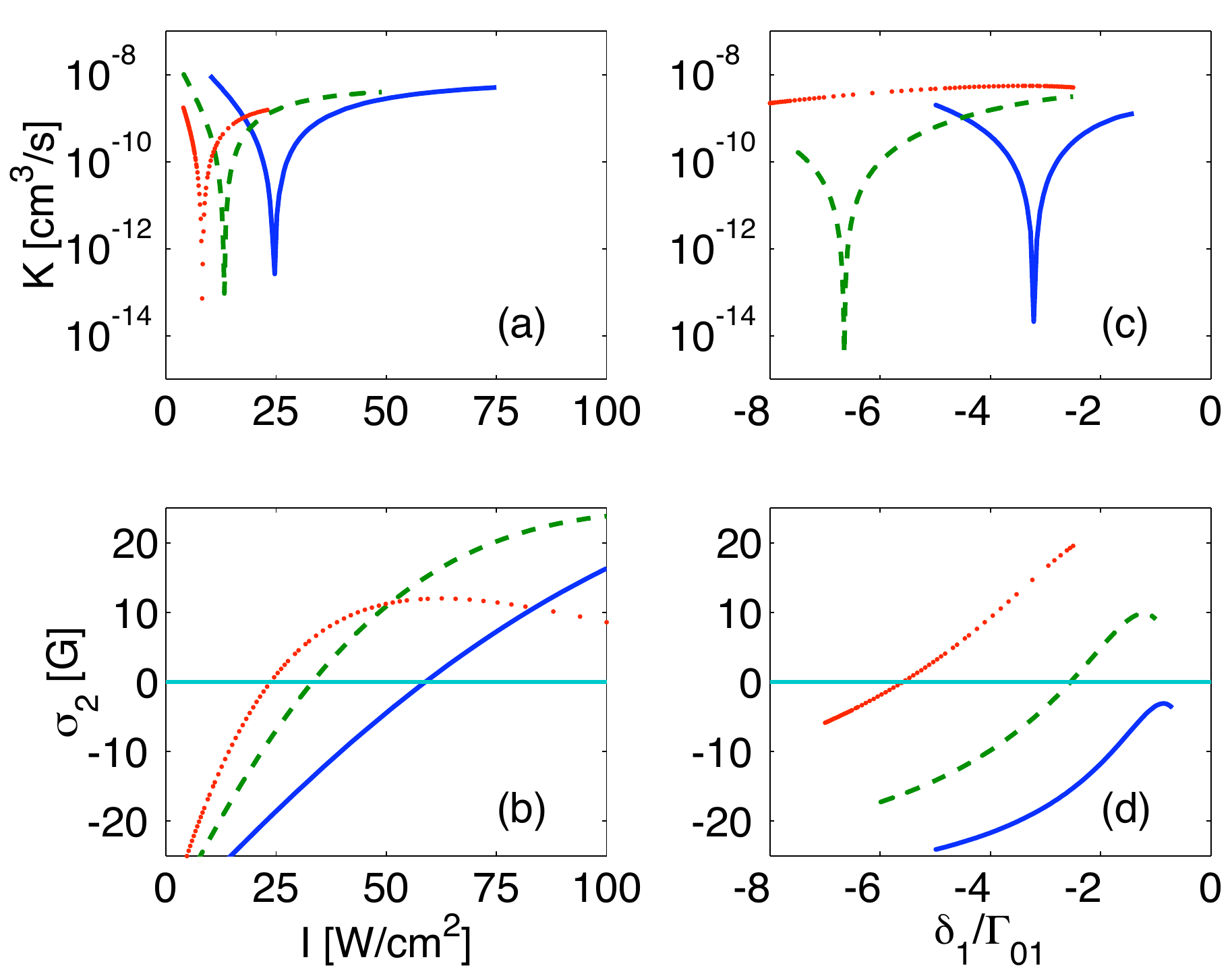}
\caption{(color online)~Rate constant and fieldshift as a function of laser intensity (a,b) and detuning (c,d), where the density is $\rho=1\times10^{12}$~cm$^{-3}$. (a,b) The blue solid (green dashed, red dotted) line is for laser detuning $-\delta_1/\Gamma_{01}=4$~(2, 1). (c,d)~ The blue solid (green dashed, red dotted) line is for laser intensity $I=20$~(40, 80)~W/cm$^2$. Note that a negative (positive) fieldshift means that magnetic field position of the Feshbach resonance decreases (increases).}
\label{MA_ALL}
\end{figure}

We also consider a resonant magnetic field and a detuned laser, in which case $i=2$ and $j=1$ in Eqs.~\eq{RATE_EQN}. For fixed laser detuning and $\Omega_1=\bar\Omega_1\sqrt{I}$, the intensity position of the node is  $I_c=(-8\pi\delta_1\omega_\rho/\eta\bar\Omega_1^2)/(\rho^{1/3}L_1)$. This node exists in addition to any nodes in the solution to the radial Schr\"odinger equation~\cite{KOS00}, which occur for arbitrary intensity. The details of the Feshbach resonance are largely irrelevant, and $I_c$ depends instead on the laser detuning and the classical size of the photassociation molecule. Specifically, for a $^{7}$Li condensate with $\rho=10^{12}$~cm$^{-3}$, as well as a red laser detuning of one linewidth, $\delta_1=-\Gamma_s$, the cessation intensity is $I_c\approx 25$~W/cm$^2$, as per Fig.~\ref{MA_ALL}~(a). Perhaps surprisingly, the rate constant actually increases as the intensity goes to zero, but this is only because magnetoassociation, the faster process, begins to dominate. Similarly, a node arises for fixed laser intensity and variable detuning, as shown in Fig.~\ref{MA_ALL}(c). Given the connection to the collision model of the previous section, the dependence of the magnetoassociation node on both laser detuning and intensity arises because these are two means for optically tuning the resonant scattering length~\cite{FED96,MAC01,THE04}.

Beyond quantum interference, the photoassociation molecule shifts the magnetic field position of the Feshbach resonance, $\sigma_2=\sigma_{02}+{\cal L}\delta_1/\Omega_3$, as shown in Fig.~\ref{MA_ALL}(b,c). Besides the dependence on both laser intensity and detuning, it appears that the nodes in the resonant magnetoassociation rate constant and fieldshift occurs at very different values of intensity/detuning, in contrast to the case of resonant photoassociation discussed in the previous section. This discrepancy is due to the difference in the sizes of the photoassociation and Feshbach molecules: if the molecular sizes were equal, the nodes would vanishes similarly; if the Feshbach molecule were taken as larger, then the magnetoassocaition nodes would match, and the photoassociation nodes would mismatch. Finally, the fieldshift node vanishes entirely if the cross-molecular shift, $\delta_1{\cal L}/\Omega_3$, never equals the background shift $\sigma_{02}$, i.e., if the optically-resonant scattering length never equals the size of the Feshbach molecule, as per Fig.~\ref{MA_ALL}(d).

\section{Conclusions}
\label{CONC}

A quantum matter optics model of combined photoassociation and Feshbach resonances in a Bose-Einstein condensate has been developed. We reported the details of an analytical model based on the weakly-bound molecule assumption, which is ultimately the same as assuming that the anomalous amplitude evolves adiabatically~\cite{MAC08,NAI08}, as well as exact numerical model that is valid even for dual resonance.  For resonant lasers and off-resonant magnetic fields, the analytical results compare well with numerical experiments that account explicitly for dissociation to noncondensate modes. This agreement is best for low density, suggesting that the weak-binding/adiabatic approximation is best applied to this regime, or that the magnetodissociation ansatz, $\gamma_2\propto\rho^{1/3}$, breaks down. The combined-resonance rate constant is larger for smaller density, and the agreement between the analytical and numerical results indicates that this density dependence is inherent to the system, and not due to the magnetodissociation ansatz.

Despite the agreement with numerical experiments on Feshbach-enhanced photoassociation, both approaches underestimate the saturated rate constant, $K_x=1.4\times10^{-7}$~cm$^3$/s, that is observed experimentally for resonant photoassociation at $B=732$~G~\cite{JUN08}. This discrepancy has two likely causes. First, the position of the node in the rate constant depends on the semi-classical size of the photoassociation molecule, $L_1=133a_0$, and we have over-estimated $L_1$ by about 25\% compared to the observed value, $L_x=106a_0$~\cite{JUN08}. A smaller value of $L_1$ would mean that $B=732$~G is further from the node, and that enhancement is therefore stronger, making for a larger rate constant.  Second, the combined rate constant increases for decreasing density, which means that the edges of the condensate, rather than the center, contribute more heartily to molecule formation in this regime. Better modeling of the molecular size, as well as full accounting of condensate inhomogeneity, could improve the agreement with observation.

Considering the rate limit for Feshbach-enhanced photoassociation, we find that the rogue and unitary limits agree only in a limited range near the Feshbach resonance. This is not terribly surprising, since unitarity does not account for quantum interference, and both models breakdown on-resonance. For $B=732$~G the two limits differ by about a factor of two, and nearer to resonance resonance the two results agree reasonably. This particular disagreement is understood as a combination of an inaccurate size for the photoassociation molecule and the assumption of a homogeneous condensate. Moreover, it makes sense that a unitary limit based on a Feshbach-resonant scattering length agrees with our results near $B$-threshold where magnetoassociation losses dominate. Lastly, our previous~\cite{JAV02} prediction for the photoassociation rate limit is $\sim6\,\omega_\rho$, which compares favorably with both the $732$~G result $\sim 8\,\omega_\rho$, and the Feshbach-resonant rate $\sim 9\,\omega_\rho$. Nevertheless, this agreement is to be taken with a grain of salt: the combined resonance rate limit saturates at the rate for magnetoassociation alone (times an interference factor), which need not be the same as the rate limit for photoassociation alone.

Finally, we discuss the result for resonant $B$-fields and off resonant lasers. In particular, we focus on the implications for many-body universality, whereby the only relevant length scale in a strongly-interacting system is the interparticle distance. In photoassociation, the existence and size of the bound molecule state is well established, even in a strongly interacting (high laser intensity) system, and its appearance in Feshbach-assisted photoassociation is therefore not terribly surprising. On the other hand, because many-body concepts dominate Feshbach resonance models, so too does universality, and the Feshbach molecule is generally neglected. Here we see that a node in the net shift of the laser-assisted magnetoassociation resonance is an important test of the validity of universal models of the Feshbach resonance. In particular, the zero-field shift is $\sigma_{01}\propto1/L_2$, which tends to infinity for point-like Feshbach molecules, and the node in the fieldshift then ceases to exist. A finite node in the fieldshift therefore indicates a lengthscale other than the interparticle distance, and a breakdown of universality.

\acknowledgements
{
This work supported by a grant from the National Science Foundation (MM, PHY-00900698), and a Diamond Research Scholarship from Temple University's Office of the Vice Provost for Undergraduate Affairs (CD).
}

\end{document}